\documentclass[a4paper,11pt]{article}
\pdfoutput=1

\usepackage{amsmath,amssymb,bm,color}
\usepackage{appendix}
\usepackage{graphicx}
\usepackage{xcolor}
\usepackage{cancel}
\usepackage{mathtools}
\usepackage{jcappub}
\usepackage[T1]{fontenc} 
\usepackage{tikz}
\usepackage{fontawesome}
\usetikzlibrary{decorations.pathmorphing}
\usetikzlibrary{decorations.pathmorphing,calc}

\newcommand{\be}{\begin{equation}}
\newcommand{\ee}{\end{equation}}
\newcommand{\bea}{\begin{eqnarray}}
\newcommand{\eea}{\end{eqnarray}}

\tikzstyle{singularity}=[line width=0.6,decorate,
                         decoration={zigzag,amplitude=2,segment length=6.17}]

\title{
Hawking radiation of Nonrelativistic Scalars: Applications to Pion and Axion Production}

\author[a]{Hao-Ran Cui,}
\author[b]{Yuhsin Tsai,}
\author[c]{Tao Xu}

\affiliation[a]{School of Physics and Astronomy, University of Minnesota, Twin Cities, MN 55455, USA}
\affiliation[b]{Department of Physics and Astronomy, University of Notre Dame, IN 46556, USA}
\affiliation[c]{Homer L. Dodge Department of Physics and Astronomy, University of Oklahoma, Norman, OK 73019, USA}

\emailAdd{cui00159@umn.edu}
\emailAdd{ytsai3@nd.edu}
\emailAdd{tao.xu@ou.edu}

\abstract{In studying secondary gamma-ray emissions from Primordial Black Holes (PBHs), the production of scalar particles like pions and axion-like particles (ALPs) via Hawking radiation is crucial. While most analyses assumed relativistic production, asteroid-mass PBHs, relevant to upcoming experiments like AMEGO-X, likely produce pions and ALPs non-relativistically when their masses exceed 10 MeV. To account for mass dependence in Hawking radiation, we revisit the greybody factors for massive scalars from Schwarzschild black holes, revealing significant mass corrections to particle production rates compared to the projected AMEGO-X sensitivity.
We highlight the importance of considering non-relativistic $\pi^0$ production in interpreting PBH gamma-ray signals, essential for determining PBH properties. Additionally, we comment on the potential suppression of pion production due to form factor effects when producing extended objects via Hawking radiation. We also provide an example code for calculating the Hawking radiation spectrum of massive scalar particles \href{https://github.com/Hao-Ran-Brook/HoRNS}{\faGithub}.}

\begin{document}

\maketitle
\section{Introduction}
\label{sec:introduction}
The concept of Primordial Black Holes (PBHs), which originate differently from standard stellar collapse, has inspired discussion in various aspects of physics beyond the Standard Model (SM). PBHs are considered viable candidates to either fully or partially account for dark matter (DM)~\cite{Carr:2020gox, Carr:2020xqk, Green:2020jor, Carr:2021bzv}. An intriguing feature of this DM candidate is that, depending on the mass of the black hole, its Hawking radiation could produce SM  particles like photons, detectable through astrophysical observations.

For PBHs with asteroid-scale masses around $M\sim 10^{15-16}$~g, they can emit Hawking radiation with temperatures $T_H\approx (10^{16}{\rm g}/M)$~MeV$\sim \mathcal{O}(1-10)$~MeV. Their relatively short lifetimes $\tau\approx 10^5(M/10^{16}{\rm g})^3$~Gyrs allow for the production of observable gamma-ray signals. These signals originate from both photons directly emitted by the PBH and secondary photons produced through the electromagnetic interaction of Hawking radiation particles, such as neutral pions that decay into a pair of photons. These gamma-ray signals can unveil the properties of PBHs, presenting exciting opportunities for observation in next-generation detectors such as AMEGO-X~\cite{Caputo:2022xpx}, e-ASTROGAM~\cite{e-ASTROGAM:2017pxr}, APT~\cite{APT:2021lhj}, COSI~\cite{Tomsick:2021wed}, GECCO~\cite{Orlando:2021get}, and MAST~\cite{Dzhatdoev:2019kay}. These upcoming experiments will span the gamma-ray signal energy range from $0.1$ to $100$ MeV, significantly enhancing signal flux sensitivity compared to the COMPTEL and Fermi-LAT experiments~\cite{Kappadath:1998PhDT, Fermi-LAT:2017opo}. By integrating gamma-ray observations with data from future gravitational wave experiments, there is potential to measure the mass spectrum of PBH and identify the primordial curvature perturbation responsible for the PBH production~\cite{Agashe:2022jgk, Xie:2023cwi}.

In addition to emitting SM particles, PBH can serve as a source for generating beyond the SM particles, even if the new particles interact with the SM sector very weakly or solely through gravity. For example, PBH can produce DM particles from their Hawking radiation~\cite{Sandick:2021gew, Allahverdi:2017sks, Bell:1998jk, Lennon:2017tqq, Gondolo:2020uqv, Bernal:2021yyb, Cheek:2021odj, Cheek:2021cfe,Bhaumik:2022zdd,Gehrman:2023esa, Kim:2023ixo, Gehrman:2023qjn} and facilitate a portal for baryogenesis~\cite{Alexander:2007gj, Baumann:2007yr, Fujita:2014hha, Hook:2014mla,Hamada:2016jnq, Morrison:2018xla, Bernal:2022pue, Datta:2020bht, Bhaumik:2022pil, Gehrman:2022imk}. PBHs can also produce axion-like particles (ALP) that subsequently decay into SM photons, adding extra contribution to gamma-ray signals~\cite{Agashe:2022phd, Jho:2022wxd}.\footnote{In this study, we focus on the ALP production from the Hawking radiation of PBHs. See~\cite{Branco:2023frw, Dent:2024yje} for the production of ALPs from the superradiance process.} The presence of a large dark sector particle population can also modify PBH's evaporation rate, providing ways to probe the existence of new particles~\cite{Ukwatta:2015iba, Baker:2021btk, Baker:2022rkn, Boluna:2023jlo}. Additionally, PBHs may be produced at the LHC~\cite{Dimopoulos:2001hw, Giddings:2001bu, CMS:2010oej}, emitting SM Higgs particles and generating distinctive signals~\cite{Nayak:2006vf,Erkoca:2009kg}. The evaporation of PBHs can also probe the lepton sector; for more details, see \cite{Lunardini:2019zob, DeRomeri:2024zqs,Dasgupta:2019cae} and references therein.

In the majority of the referenced literature, Hawking radiation is typically estimated by assuming that the emitted particles have rest masses comparable to or smaller than the Hawking temperature (e.g.,~\cite{PhysRevD.13.198, PhysRevD.14.3260}). Consequently, the Hawking radiation particles are moving at relativistic speeds after being produced. When the particle mass is non-negligible relative to the Hawking temperature, two main effects need to be included for a more accurate estimation of the emission rate. Firstly, massive particles are produced on-shell from the PBHs, meaning their energy must be greater than their rest mass for an observer located at infinity. Secondly, the absorption rate of massive particles differs from that of massless particles. The effect of absorption near the horizon is embedded in the so-called greybody factor. These effects become particularly significant in the context of massive particles with masses exceeding the Hawking temperature, where production is most efficient in the non-relativistic scenario due to the nature of black hole thermodynamics. In some studies, non-relativistic corrections to the production of massive particles are speculated, with the effect of the particle mass considered in the kinematic condition of the energy spectrum, but without taking into account the exact greybody factors of massive particles~\cite{Arbey:2019mbc, Arbey:2021mbl}. 

While this relativistic approximation in the greybody factor may suffice for order-of-magnitude estimations when particles are predominantly produced relativistically, precise energy spectrum determination—such as for gamma-rays from SM pions or ALPs with masses comparable to the Hawking temperature—requires careful consideration of non-relativistic corrections to the emission rate of massive particles. The Hawking radiation of massive fermions has been discussed in literatures~\cite{PhysRevD.14.3251, Doran:2005vm, Dolan:2006vj} with the application to the production of fermionic DM~\cite{Cheek:2021odj}. Compared to this, although the study of massive scalar radiation exists~\cite{ Jung:2004yh,Castineiras:2007ma,Benone:2014qaa, Benone:2017xmg,osti_7131626,PhysRevD.14.3251} and has been used to estimate the SM pion production~\cite{PhysRevD.41.3052,1991ApJ...371..447M,1991PhRvD44376M}, the result has not been applied more generally to other scalar particle production, such as the ALP. In sight of future prospects for precise Hawking radiation measurements with upcoming ${\rm MeV}$ gamma-ray telescopes, the accurate calculation of PBH Hawking radiation spectra, incorporating all non-relativistic effects, becomes increasingly important.

In this paper, we revisit the computation of Hawking radiation of massive and charge-neutral scalar particles, showing the significance of non-relativistic corrections in pion or ALP production when comparing resulting gamma-ray spectra to experimental sensitivities. Alongside presenting results for benchmark model examples in the paper, we offer a code for calculating the Hawking radiation of massive scalar particles. Our study examines the emission of neutral scalar particles from non-rotating, chargeless black holes. However, the calculation can be extended to black holes with charge and spin by employing a more general equation for the scalar particle's wave function.

The outline of the paper is as follows. In Sec.~\ref{sec:HawkingRadiation}, we review the derivation of Hawking radiation for massive scalar particles and contrast our findings with existing literature, which relies on assumptions of massless particles or simplifications in energy dependence to a non-relativistic form. In Sec.~\ref{sec:gammaray}, we calculate the gamma-ray spectrum from PBH's Hawking radiation, including contributions from the non-relativistic pion or ALP production and decay. In Sec.~\ref{sec:indirectdetection}, we compare the gamma-ray spectrum with and without proper calculation of non-relativistic pion or ALP production and show that the future AMEGO-X measurement can be sensitive to the improper treatment of the scalar production. We conclude in Sec.~\ref{sec:discussion}. 

\section{Hawking Radiation Rate of Massive Scalar Particles}
\label{sec:HawkingRadiation}

In this section, we discuss the calculation of the Hawking radiation rate of massive scalar particles from a Schwarzschild PBH. We use natural unit and set $\hbar=c=k_B=1$ in this work.

Black holes possess thermal properties and can emit Hawking radiation~\cite{Hawking1975} with energy spectrum similar to the blackbody radiation~\cite{Page1976}
\bea
\frac{dN_{\omega l n s q}}{dtd\omega}=\frac{\langle N_{\omega l n s q}\rangle }{2\pi}=\frac{1}{2\pi}\frac{\Gamma_{\omega l n s q}}{e^{\omega/T_H}+ (-1)^{2s+1}}\,,
\label{eq:HawkingRadiation}
\eea
where $\langle N_{\omega l n s q}\rangle $ is the expected number of the particles with energy $\omega$, angular momentum $l$, azimuthal quantum number $n$, spin $s$, and electric charge $q$ observed at infinity, 
and $T_H=(8\pi G M)^{-1}$ is the black hole temperature with black hole mass $M$. The greybody factor $\Gamma_{\omega l n s q}$ is the correction to a black body radiation spectrum in flat spacetime and is dependent on the properties of emitted particles and the black hole.

The curved spacetime outside the black hole's event horizon effectively acts as a potential barrier. Consequently, only a fraction of particles radiated by the black hole can penetrate the barrier and reach a distant particle detector, with the transition rate being the greybody factor $\Gamma_{\omega l n s q}$. The remaining fraction $1-\Gamma_{\omega l n s q}$ of the particles is scattered back to the black hole. For a more detailed description of the scattering process based on the Penrose diagram, see Appendix~\ref{Penrose}.

In this study, we model PBHs as Schwarzschild black holes. This assumption is based on the expectation of rapid PBH spin loss due to the superradiance process and rapid PBH charge loss due to Hawking radiation. The superradiance instability is triggered when the wavelength of the scalar particle is comparable to the horizon size of the PBH~\cite{PhysRevD.22.2323, Furuhashi2004, Dolan2007}. This superradiance condition is naturally satisfied in the scenario where non-relativistic Hawking radiation can be important, $m\sim \mathcal{O}(1/G M)$, thus efficiently eliminating the PBH spin in the benchmarks of this study. Regarding the PBH charge loss rate, it is found that charged non-rotating black holes in the mass range we consider can Hawking radiate away their charges quickly~\cite{Hiscock:1990ex}. Therefore, we focus on the Hawking radiation from Schwarzschild PBHs. Additionally, we assume that PBHs maintain constant masses in the analysis since the lifetime of the PBHs is much longer than the observation time of the indirect detection experiments.

\subsection{Greybody Factor}
The correction from the particle mass to the radiation is encoded in the greybody factor, therefore in this subsection, we review the calculation of $\Gamma_{\omega l n s q}$ for charge-neutral massive scalar particles $(q=s=0)$, following the discussion in~\cite{Carolina2014}, and shall omit the subscripts $s$ and $q$ in $\Gamma_{\omega l n s q}$ for simplicity.
 
Firstly, we define the greybody factor by the Hawking radiation process with reversed time~\cite{Hawking1975}, where the particles propagate backward from infinity to the black hole. Assuming $N$ particles at infinity initially ($r\to\infty, t\to +\infty$), then $N\times \Gamma_{\omega l n}$ particles will penetrate the potential barrier and $N\times (1-\Gamma_{\omega l n})$ particles will be scattered back to infinity ($r\to\infty$, $t\to -\infty$), so the greybody factor is
\begin{equation}
\Gamma_{\omega l n}=1-\frac{N\times (1-\Gamma_{\omega l n})}{N}=1-\frac{\text{number of particles at $r\to\infty, t\to -\infty$}}{\text{number of particles at $r\to\infty, t\to +\infty$}},
\label{eq: def greybody}
\end{equation}
which only requires the information of the particles at $r\to \infty$.\footnote{The time-reversed process introduced here corresponds to a different boundary condition from the standard Hawking radiation, which is more difficult for computational simulation.} 
Secondly, we study the behavior of particles from Hawking radiation by treating them as classical fields~\cite{Mukhanov:2007zz,DeWitt:1975ys}. The number of particles is proportional to the square of the amplitude of the incoming ($r\to\infty, t\to +\infty$) and outgoing ($r\to\infty, t\to -\infty$) scalar field.

The equation of motion in curved spacetime of charge-neutral massive scalar field $\Psi$ such as neutral pion $\pi^0$ and ALP is
\begin{align}
\nabla_{\nu}\nabla^{\nu}\Psi={m}^2\Psi,
\label{field eq}
\end{align}
where $\nabla_{\nu}$ is the covariant derivative and $m$ is the field mass. The following Schwarzschild line element is used,
\begin{align}
    ds^2=f(r) dt^2- \frac{1}{f(r)}dr^2-r^2d\theta^2-r^2\text{sin}\theta^2 d\phi^2,
\end{align}
where $f(r)=1-r_s/r$ and $r_s=2GM$ is the horizon radius with $M$ being the black hole mass. 

\begin{figure}[t]
    \centering
\includegraphics[scale=1.2]{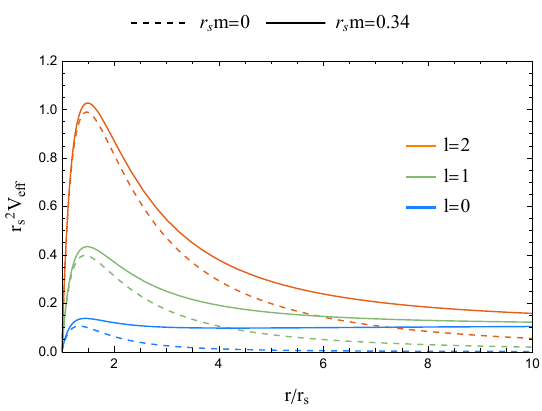}
    \caption{The effective potential $V_{\rm eff}$ normalized by $r_s^2$ for different particle masses and angular momenta in Schwarzschild spacetime. As $r_s\to \infty$, the potential approaches $r_s^2 m^2$. The solid lines of $r_s m=0.34$ may represent the potential for $\pi^0$ with $m=135$ MeV around the Schwarzschild black hole of $10^{14.5}$ g. Note that the larger the angular momentum, the higher the peak of the potential, and hence the lower the transition rate (with the same energy). Therefore, s-wave dominates the radiation in the non-relativistic limit. 
    }
    \label{effv}
\end{figure}

By separating the variables in spherical coordinates, the mode with energy $\omega$ propagating backward in time can be expanded as~\cite{Futterman1988} 
\begin{align} \Psi(r,\theta,\phi,\omega,M)=\sum_{n, l}e^{+i\omega t}\frac{R(l, r,\omega, M)}{r}K_{ln}Y_{ln}(\theta,\phi) ,
\label{monochromatic solution}
\end{align}
where $R(l, r,\omega, M)$ is the radial function, $Y_{ln}(\theta, \phi)$ is the spherical harmonics with orbital quantum number $l$ and azimuthal quantum number $n\in [-l,l]$, and $K_{ln}$ is the combination coefficient.\footnote{Note that the frequency $\omega$ is defined by $\omega \Psi=-i (\partial/\partial t)\Psi$, where $\partial/ \partial t$ is the time-like killing vector. This frequency can be identified with the energy $E$ only at infinity where $f(r)\to 1$ and the coordinate time $t$ approaches the proper time $\tau$ of observers. However, $\omega$ is still the most convenient parameter to label different modes of the field since it is always constant due to the property of Killing vectors.} The radial function $R(l, r,\omega, M)$ satisfies 
\begin{align}
\left[-\frac{d^2}{d{r^*}^2} +V_{\rm eff}(l,r)-\omega^2\right] R(l, r,\omega, M)=0,
\label{radial eq}
\end{align}
where $r^*$ is the tortoise coordinate defined by $\frac{d}{d r^*}=f(r)\frac{d}{d r}$, and the effective potential is\footnote{Notice that in the Eddington-Finkelstein coordinate, the effective mass $\sqrt{f(r)}m\to 0$ when $r\to r_s$ so that the particle's world-line can exist in the narrow light cone near the horizon.} 
\begin{align}
    V_{\rm eff}(l,r)=f(r)\left[m^2+\frac{l(l+1)}{r^2}+\frac{2GM}{r^3}\right],
    \label{veff}
\end{align}
which is plotted in Fig.~\ref{effv} with different masses and angular momentum $l$. Note that when $M=0$, $V_{\rm eff}$ is the regular potential of a scalar field, and $f(r)$ in the front and the last term are the corrections from curved spacetime. 

We numerically solve Eq.~(\ref{radial eq}) and match the result from the viewpoint of quantum mechanical scattering theory, where the real scattering solution $R(l, r,\omega, M)$ should be only incoming wave at the near-horizon region ($r\to r_s$) and a mixture of incoming and outgoing waves at the asymptotic region ($r\to \infty$) as following
\begin{align}\label{eq.R}
R(l, r,\omega, M)=
\left\{
\begin{array}{cc}
   \sqrt{v}\,T_{\omega l} e^{-i\omega r^*}  &  (r\to r_s)\,,\\
     e^{-i\omega v r^*}+ R_{\omega l} e^{i\omega v r^*} & (r\to \infty)\,.
\end{array}
    \right.
\end{align}
Here $v=\sqrt{1-m^2/\omega^2}$ is the velocity of the scalar particle. We normalize the amplitude of the incoming mode exp$(-i\omega v r^{*})$ to unity for the particle propagating back from infinity. Another incoming mode exp$(-i\omega r^{*})$ describes the particle that penetrates the gravitational barrier and enters the black hole with the transition amplitude $T_{\omega l}$. The outgoing mode exp$(i\omega v r^*)$ corresponds to the particle scattered back to infinity with the reflection amplitude $R_{\omega l}$. The transition and reflection amplitudes are independent of $n$ in the spherically symmetric spacetime. In the numerical simulation, $T_{\omega l}$ and $R_{\omega l}$ are calculated by comparing the amplitudes in front of each mode after integrating the radial equation from $r_s=2GM$ to a large distance~\cite{Park2004, Persides1973, Sanchez1977} (practically more than 1000$\times r_s$). More discussion on $T_{\omega l}$ and $R_{\omega l}$ can be found in Appendix~\ref{amplitude justification}. 

\begin{figure}[ht]
\centering
\includegraphics[scale=1.1]{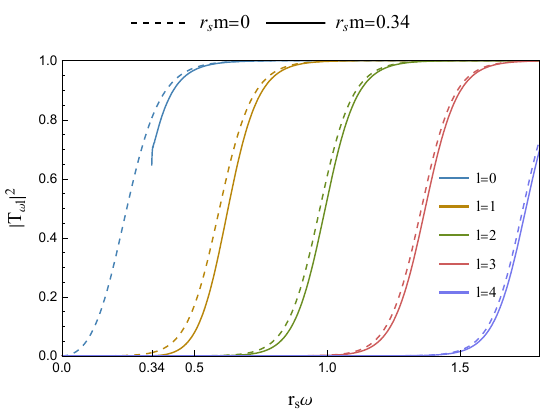}\\\vspace{1em}
\includegraphics[scale=1.1]{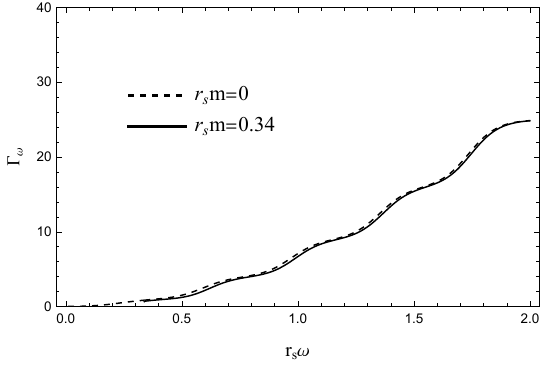}
\caption{\emph{Upper panel}: The transition rate $|T_{\omega l}|^2$ of massless and massive scalar particles with different angular momenta. \emph{Lower panel}: The greybody factor $\Gamma_{\omega}=\sum_{l}(2l+1)|T_{\omega l}|^2$ of massless and massive scalars. Note that there is a hard cutoff at $\omega=m$ ($r_s\omega=r_s m=0.34$) for the massive particle. As $\omega$ increases, the modes of higher $l$ are excited, hence the bumps in the lower panel.
Note that $r_s\omega=\omega/4\pi T_H$ for Schwarzschild black holes. } 
\label{transitionrate}
\end{figure}

In Fig.~\ref{transitionrate}, we show $|T_{\omega l}|^2$ with different $l$ and $\omega$ for different choices of particle masses. If we treat $\omega^2$ as an effective energy, then Eq.~(\ref{radial eq}) is a time-independent Klein-Gordon equation. When $\omega^2<V_{\rm eff}$, the transition rate comes from a quantum tunneling process, and therefore $|T_{\omega l}|^2\to0$ when $\omega \to0$. In other words, the particle's wavelength goes to infinity when $\omega\to0$, thus the black hole appears as a mere point to the particle, suppressing the possibility of transition into the black hole. When $\omega^2$ exceeds the maximum value of $V_{\rm eff}$, the wave function describes a regular scattering process where $|T_{\omega l}|^2\to1$ as $\omega$ increases. We see both behaviors in the plot for each $l$-mode. Based on the definition of the greybody factor Eq.~(\ref{eq: def greybody}), we have
\[
\Gamma_{\omega l n}=|T_{\omega l}|^2, \hspace{3mm} n\in[-l, l].
\]
Given that the spacetime is spherically symmetric, we sum over the contributions from the modes with different angular momenta to study the energy dependence of the greybody factor, 
\begin{align}
\Gamma_{\omega}=\sum_l\sum_{n=-l}^l \Gamma_{\omega l n}=\sum_l (2l+1)|T_{\omega l}|^2, 
\end{align}
Note that as the transition rate of a certain mode, $\Gamma_{\omega l n s}<1$, but $\Gamma_{\omega}$ can be larger than 1. 

\begin{figure}[ht]
    \centering
\includegraphics[scale=1.1]{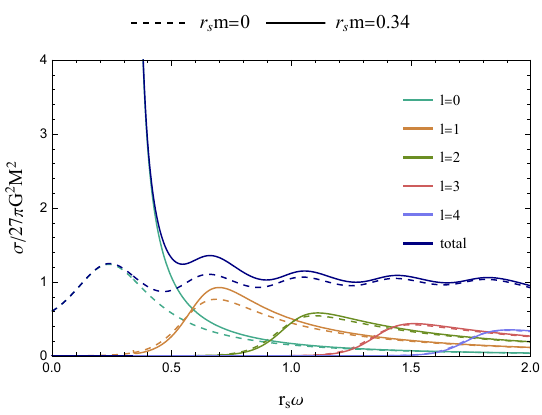}
    \caption{The normalized cross section $\sigma/27\pi G^2M^2$ of massless (dashed lines) and massive scalars (solid lines). In the low-energy limit, the cross section of the massive scalar blows up, and the massless approaches the black hole area $16\pi G^2M^2$. In the high-energy limit, the massive and the massless cross sections both approach the geometrical optics limit $27\pi G^2M^2$.}
    \label{Fig.sigma}
\end{figure}

\begin{figure}[t]
    \centering
\includegraphics[scale=1.13]{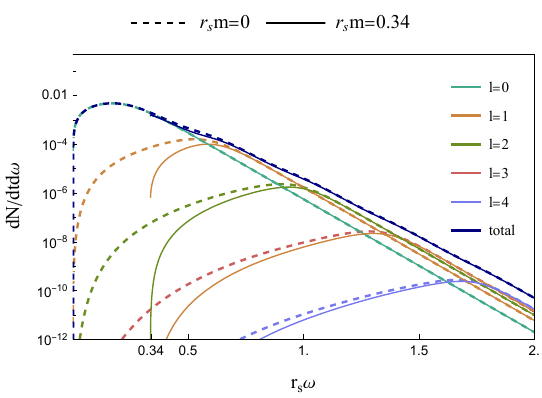}
    \caption{The particle production rate (log-plot) measured at infinity for massless and massive scalars. The production rates of different modes of different $l$ are compared. The dark blue line shows the total production rate when summing over all contributions of angular momentum, and there is a cut-off at $\omega=m$ ($r_s\omega=r_s m=0.34$) for the massive particle. The production rate of massless particles peaks at $r_s \omega\approx 0.2$, which in the simple case of Schwarzschild black holes means $\omega/T_H=4\pi\times0.2\approx 2.5$.}
    \label{dnrate}
\end{figure}

When considering the Hawking radiation process with reversed time, the absorption cross section of the radiation for massive particles is~\cite{Unruh1976, Carolina2014}
\begin{align}
\sigma_{\rm massive}=\frac{\pi}{(\omega v)^2} \sum_{l}(2l+1)|T_{\omega l}|^2=\frac{\pi}{(\omega v)^2} \Gamma_{\omega}\,.
\label{eq:crosssectionandgamma}
\end{align} 
After summing over all contributions of angular momenta, the energy spectrum of produced scalar particles relates to the cross section as~\cite{Page1977}
\bea
    \frac{dN}{dtd\omega }=\frac{1}{2\pi}\frac{27 G^2M^2\omega^2 v^2 }{e^{\omega/T_H}-1}\left(\frac{\sigma_{\rm massive}}{27\pi G^2M^2}\right)=\frac{1}{2\pi}\frac{ \Gamma_{\omega}}{e^{\omega/T_H}-1}\,,
\label{eq:NRcalculation}
\eea
where $27\pi G^2M^2$ is the cross section in the geometric optics limit~\cite{Andersson2000} (with radius $\frac{\sqrt{27}}{2}r_s$ of black hole shadow).\footnote{Some works also call $\left(\frac{\sigma}{27\pi G^2M^2}\right)$ the greybody factor~\cite{Aharony1999}.} We denote results calculated with Eq.~\eqref{eq:NRcalculation} for the non-relativistic particle production as ``NR calculation'' in the rest of the paper. We provide a package for the calculation of the cross section and the scalar Hawking radiation rate \href{https://github.com/Hao-Ran-Brook/HoRNS}{\faGithub}. In Fig.~\ref{Fig.sigma} and~\ref{dnrate}, we show examples of the cross section and particle production rate between massless and massive scalars with two black hole masses. 

\subsection{Comparison to Previous Estimates}
As shown in Figs.~\ref{Fig.sigma} and~\ref{dnrate}, the greybody factor and particle production rate differ significantly between massless (dashed) and massive (solid) scalars in the non-relativistic regime $\omega\to m$ and $v<1$. Although $|T_{\omega l}|^2$ remains finite as $\omega\to m$, $\sigma$ becomes infinity due to the denominator $(\omega v)^2=\omega^2-m^2\to 0$ in Eq.~\eqref{eq:crosssectionandgamma}. The idea is that when a particle is massive and highly non-relativistic, the black hole can always trap the particle. The signal production rate also decreases quickly when the energy is lowered to the scalar mass threshold.

Hawking radiation has been studied for producing massive scalars such as pions, Higgs, and ALPs~\cite{Agashe:2022phd, Jho:2022wxd}. However, most previous analyses approximate the production rate by neglecting the particle mass or by making rough estimates of non-relativistic corrections (see, however,~\cite{PhysRevD.14.3251,Doran:2005vm,Dolan:2006vj} for an explicit treatment of massive pions in the gamma-ray calculation). It is therefore essential to compare the more accurate particle production rates with earlier approximations in order to assess whether the differences are significant relative to experimental sensitivities.

\begin{figure}[t]
    \centering
    \includegraphics[scale=1.1]{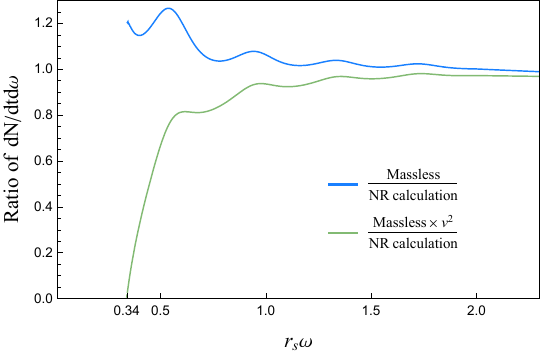}
    \caption{The ratio of the production rates of $\pi^0$ calculated by different methods with $M=10^{14.5}g$ and $r_s m=0.34$. The blue curve is the ratio $\left(\frac{dN}{dtd\omega}\right)_{\rm massless}/\left(\frac{dN}{dtd\omega}\right)_{\rm NR\hspace{1mm} calculation}=\sigma_{\rm massless}/v^2\sigma_{\rm massive}$, namely the deviation of approximating the massive greybody factor by the massless. The green curve shows $\left(\frac{dN}{dtd\omega}\right)_{\rm Massless\times v^2}/\left(\frac{dN}{dtd\omega}\right)_{\rm NR \hspace{1mm} calculation}=\sigma_{\rm massless}/\sigma_{\rm massive}$, which indicates the validity of the approximation Eq.~\eqref{eq:vsqmassless}. In the high-energy limit, all curves converge to 1 as indicated in Fig.~\ref{Fig.sigma} that $\sigma_{\rm massless} \to \sigma_{\rm massive}$ in the high-energy limit and meanwhile $v^2=1-m^2/\omega^2\to 1$. However, they differ at the non-relativistic limit, where the massless calculation gives roughly an extra $20\%$ of the photon numbers under the current parameters. The additional $v^2$ factor in the Massless$\times v^2$ method suppresses the production rate in the non-relativistic limit.}
    \label{productionratecompare}
\end{figure}

When ignoring the particle mass, the production rate is written as
\bea
\left(\frac{dN}{dtd\omega}\right)_{\rm Massless}=&\frac{1}{2\pi}\frac{27 G^2M^2\omega^2 }{e^{\omega/T_H}-1}\left(\frac{\sigma_{\rm massless}}{27\pi G^2M^2}\right)=\frac{1}{2\pi}\frac{ \Gamma_{\rm massless}}{e^{\omega/T_H}-1}\,. \label{eq:massless}
\eea
Here $\Gamma_{\rm massless}$ denotes the greybody factor for massless scalars, and $\sigma_{\rm massless}$ is calculated following the same procedure but with particle mass $m=0$. We take the numerical values of $\Gamma_{\rm massless}$ implemented in the package BlackHawk~\cite{Arbey:2019mbc, Arbey:2021mbl}. Instead of properly including the scalar mass, some literature includes the non-relativistic correction by keeping the velocity $v^2$ term while using the massless cross section~\cite{Cheek:2021odj, Cheek:2022dbx, Cheek:2022mmy}
\bea
    \left(\frac{dN}{dtd\omega}\right)_{{\rm Massless}\times v^2}=&\frac{1}{2\pi}\frac{27 G^2M^2\omega^2 v^2}{e^{\omega/T_H}-1}\left(\frac{\sigma_{\rm massless}}{27\pi G^2M^2}\right)=\frac{1}{2\pi}\frac{v^2 \Gamma_{\rm massless}}{e^{\omega/T_H}-1}.
\label{eq:vsqmassless}
\eea
Eqs.~\eqref{eq:NRcalculation}, \eqref{eq:massless}, and \eqref{eq:vsqmassless} agree with each other in the relativistic limit $v\to 1$ but differ for the non-relativistic production ($\omega\approx m$), as is shown in Fig.~\ref{productionratecompare}. Throughout this study, results obtained using Eq.~\eqref{eq:massless} are referred to as "Massless", and results obtained using Eq.~\eqref{eq:vsqmassless} are referred to as "Massless$\times v^2$".

\subsection{Possible Suppression for Composite Particles}\label{sec.tidle}

The Hawking radiation of composite particles is studied in~\cite{Johnson:2018gjr}, and the production rate is found to be suppressed when the tidal force effect on spatially extended objects is included. This could result in an effective reduction of the greybody factor for composite particles, such as pions being QCD bound states, in the low momentum regime. To estimate the suppression in the production rate of composite particles, one must compare the particle's physical radius to the PBH horizon radius. Generally, the production rate is exponentially suppressed when the particle radius is much larger than the Schwarzschild radius. Conversely, Hawking radiation rate is also suppressed by the particle mass when its radius is much smaller than the Schwarzschild radius. Thus, accurately determining the pion radius is crucial for incorporating this suppression.

The definition of the pion radius varies based on how its properties are probed. For instance, gravitational form factor studies suggest a pion light cone mass radius of around $0.3$~fm~\cite{Freese:2019bhb, Choudhary:2022den, Xu:2023izo}. Charge radius measurements indicate a radius of approximately $\sqrt{\langle r_\pi^2\rangle}\simeq 0.8~{\rm fm}$~\cite{Workman:2022ynf}, while the Compton wavelength of the neutral pion is about 9 fm. Given the limited understanding of the hadronization process near the PBH horizon, the correct radius for Hawking radiation calculations remains unclear. If we apply the suppression factor $\sim\exp(-\frac{s}{3.3})$ as discussed in~\cite{Johnson:2018gjr}, where $s\simeq \frac{r_\pi^3}{(G\,M)^3}$, the suppression ranges from $\mathcal{O}(0.1)$ (for 0.3 fm) to $10^{-\mathcal{O}(7000)}$ (for 9 fm) for a $10^{14.5}$~g black hole.

Given the large uncertainty in form factor calculations, we focus on clarifying the non-relativistic effects in this study by treating pions as fundamental particles, similar to ALPs, in their production via Hawking radiation.

\section{Gamma-ray Spectrum from Hawking Radiation} 
\label{sec:gammaray}
One important channel to examine the precise calculation of Hawking radiation spectrum is the gamma-ray searches for PBHs. 
The precise calculation of $\Gamma_{\rm massive}$ is important for PBHs within the asteroid mass range that have Hawking temperature $T_H \simeq 0.1$-$10~{\rm MeV}$. These PBHs are stable enough to produce gamma-ray signals in the present universe. The PBH-produced SM pions and ALPs of mass from $10$ to $100$~MeV contain non-negligible fraction of non-relativistic contributions. The proper inclusion of the non-relativistic correction is therefore important for predicting the correct photon spectrum in their subsequent decay into photons.

The visible particles emitted in the Hawking radiation process can be used to discover the existence of PBHs and further test the Hawking radiation mechanism. There are several studies on using indirect detection experiments to detect gamma-ray signals from PBH evaporation in the local galaxy \cite{Laha:2019ssq,DeRocco:2019fjq,Laha:2020ivk, Coogan:2020tuf, Ray:2021mxu} as well as dwarf galaxies \cite{Coogan:2020tuf} and the intergalactic medium \cite{Ray:2021mxu}. Gamma-ray observations pointing at the Galactic Center~(GC) has several advantages. First, the higher DM abundance at the GC leads to a stronger gamma-ray flux. Second, the GC gamma-rays are observed with negligible redshift and propagation effects, allowing a better reconstruction of the original spectrum shape for the study of the Hawking radiation rate. Therefore, we focus on the GC indirect detection signal in this work.

To calculate the flux of the galactic gamma-ray signal, we start with the photon production rate from a single Schwarzschild PBH following \cite{Agashe:2022jgk}. The total photon flux contains contributions from primary photons directly emitted by the PBH, and secondary photons produced by the electromagnetic interaction of Hawking radiation products. We assume the secondary photons are generated soon after the Hawking radiation such that we the direction of the secondary photon is approximated with the direction of the PBH. The primary photon emission rate is given in Eq.~\eqref{eq:HawkingRadiation},
\bea
\frac{dN_{\gamma, {\rm primary}}}{d E_{\gamma}dt}=\frac{1}{2\pi}\frac{\Gamma_\gamma}{e^{E_\gamma/T_H}-1}.
\label{eq:primary}
\eea
Here $\Gamma_\gamma$ is the greybody factor for vector particles. We take the photon emission greybody factor from BlackHawk~\cite{Arbey:2019mbc, Arbey:2021mbl}. 

The secondary photon flux is from two sources, the particle decay and the final state radiation~(FSR). The former is from diphoton decay of neutral particles produced by Hawking radiation. The latter contribution is from the radiation of charged particles from the PBH. The photon energy spectrum in the diphoton decay process of the mother particle mass $m_i$ and energy $\omega_i$ is
\bea
    \frac{d N_{i,\textrm{decay}}}{d E_\gamma} &=& \frac{\Theta(E_\gamma - E_i^-) \Theta(E_i^+ - E_\gamma)}{E_i^+ - E_i^-}\,,\\
    E_i^\pm &=& \frac{1}{2} \left ( \omega_i \pm \sqrt{\omega_i^2 - m_i^2} \right )\,.
\eea
Note we use $\omega_i$ specifically for the energy of scalar particles produced by Hawking radiation to match notations used in Sec.~\ref{sec:HawkingRadiation}. The decay spectrum is used to calculate photons from neutral pions in the SM scenario and ALPs for new physics searches. The decay spectrum is incorporated with the detailed calculation of scalar particle emission spectra from PBHs in Sec.~\ref{sec:HawkingRadiation}, giving the final photon spectrum in the decay component,
\bea
\frac{d N_{\gamma,\textrm{decay}}}{d E_\gamma d t} = \int d \omega_i \, 2\frac{d N_{i}}{d \omega_i d t} \frac{d N_{i,\textrm{decay}}}{d E_\gamma}.
\label{eq:decay}
\eea

Another source of secondary photon is FSR where photon spectrum is calculated from a charged particle of mass $m_i$ and energy $E_i$ \cite{Chen:2016wkt, Coogan:2019qpu},
\bea
    \frac{d N_{i,\textrm{FSR}}}{d E_\gamma} &=& \frac{\alpha}{\pi Q_i}P_{i\rightarrow i\gamma}(x) \left [\log \left (\frac{1-x}{m_i^2} \right ) -1 \right ]\,,\\
    P_{i\rightarrow i\gamma}(x) &=& \begin{dcases} \frac{2(1-x)}{x}, & i=\pi^\pm \\ \frac{1+(1-x)^2}{x}, & i=\mu^\pm, e^\pm \end{dcases}.
\eea
here $x\equiv2E_\gamma/Q_i$, $\mu_i\equiv m_i/Q_i$ and we choose the FSR energy scale $Q_i=2E_i$ to be twice of the energy of charged particles. This gives the FSR contribution to the total gamma-ray flux,
\bea
\frac{d N_{\gamma,\textrm{FSR}}}{d E_\gamma d t} = \int d E_i \frac{d N_{i}}{d E_i d t} \frac{d N_{i,\textrm{FSR}}}{dE_\gamma}\,.
\label{eq:FSR}
\eea
With Eqs.~\eqref{eq:primary}, \eqref{eq:decay}, and \eqref{eq:FSR}, we obtain the total photon spectrum from a single PBH,
\bea
    \frac{d N_{\gamma,\textrm{tot}}}{d E_\gamma d t} &=& \frac{d N_{\gamma,\textrm{primary}}}{d E_\gamma d t} + \sum_{i=\pi^0,(a)} \frac{d N_{\gamma,\textrm{decay}}}{d E_\gamma d t} + \sum_{i=e^\pm,\mu^\pm,\pi^\pm} \frac{d N_{\gamma,\textrm{FSR}}}{d E_\gamma d t}.
\label{eq:HawkingRadiationTotal}
\eea

We will use Eq.~\eqref{eq:HawkingRadiationTotal} in the signal analysis for distinguishing the non-relativistic effect of scalar particle emission. In particular, we take two example scalar particles, the neutral pion $\pi^0$ and the ALP $a$, to show the implication of this study for future gamma-ray searches. In the neutral pion case, we only include the pion emission in the second term in Eq.~\eqref{eq:HawkingRadiationTotal}. The neutral pion production rates are calculated with methods discussed in Sec.~\ref{sec:HawkingRadiation} and show the implications for indirect detection searches in the next section. In the ALP case, we include both the neutral pion emission and the ALP emission in the decay contribution of secondary photons. We keep using the Massless method to calculate the neutral pion emission rate, while varying different methods to obtain the ALP production rate. We checked the pion production is sub-dominant with our benchmark PBH and ALP masses, and the gamma-ray spectrum feature is mostly determined by the ALP decay. We also use the Massless method for the Hawking radiation rate of charged pions $\pi^{\pm}$ to calculate their FSR photon flux, in both the pion case and the ALP case. The FSR from charged pions is subdominant due to the pion mass. The final results are shown in the following section. 

\section{Application to Indirect Detection Searches}
\label{sec:indirectdetection}
We apply the gamma-ray spectrum calculation to the searches for asteroid-mass PBHs in indirect detection experiments. The PBHs make up a fraction $f_{\rm PBH}$ of the DM relic abundance with a monochromatic mass value $M$, and we assume their spatial distribution tracks the Milky Way DM density profile. The photon flux from each PBH is isotropic. The gamma-ray flux at the earth location is the sum of photon flux from all PBHs and can then be averaged over the $4\pi$ sphere,
\bea
    \frac{d \Phi_\gamma}{d E_\gamma} = J_D \frac{\Delta \Omega}{4 \pi} \frac{f_\textrm{PBH}}{M} \frac{d N_{\gamma,\textrm{tot}}}{d E_\gamma d t}.
    \label{eq:flux}
\eea
Here $J_D$ is the same J-factor of decaying DM calculated by integrating the DM density along the line-of-sight (LOS), $dl$, within the observation solid angle $\Delta\Omega$.
\bea
    J_D = \frac{1}{\Delta \Omega}\int_{\Delta \Omega} d\Omega \int_\textrm{LOS} dl \rho_\textrm{DM}.
\eea
We take an NFW profile \cite{Navarro:1995iw} for the $\rho_{\rm DM}$ with halo parameters $r_s=11~{\rm kpc}$, $\rho_s=0.838~{\rm GeV}/{\rm cm}^3$, $r_{200}=193~{\rm kpc}$, and $r_\odot=8.122~{\rm kpc}$~\cite{2019JCAP...10..037D} for the galactic DM distribution $\rho_{\rm DM}$. We use $J_D = 1.597 \times 10^{26}~{\rm MeV} {\rm cm}^{-2} {\rm sr}^{-1}$ for our assumption of an observation region of $|R|< 5^{\circ}$ from the Galactic Center, corresponding to $\Delta{\Omega} = 2.39\times 10^{-2}~{\rm sr}$. The benchmark values of $f_{\rm PBH}$, defined as the fraction of DM energy density in the form of PBH, are choose to be below the existing constraints adapted from~\cite{Essig:2013goa, Fermi-LAT:2017opo} and at the same time enable the signal to be detected by future MeV scale indirect detection experiments. We use AMEGO-X \cite{Fleischhack:2021mhc,Caputo:2022xpx} as an example to show the non-relativistic effect in future observations.

\begin{figure}[t!]
\centering
\includegraphics[scale=1.3]{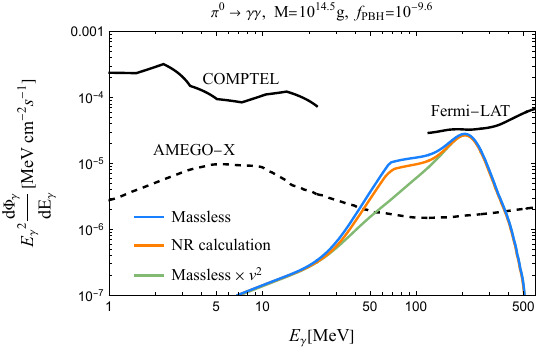}
\\\vspace{1em}
\includegraphics[scale=1.3]{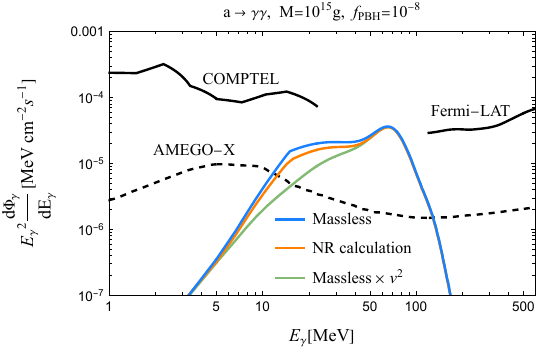}
\caption{Gamma-ray spectrum from Hawking radiation of PBHs. {\it Upper panel:} comparing different treatments of neutral pion production from a $M = 10^{14.5} {\rm g}$ PBH. {\it Lower panel:} comparing different treatments of $m_a=30~{\rm MeV}$ ALP production from a $M = 10^{15} {\rm g}$ PBH. Here we assume the tidal force effect is negligible for the pion production.}
\label{fig:GammaraySpectrum}
\end{figure}

In Fig.~\ref{fig:GammaraySpectrum}, we show the instantaneous gamma-ray spectrum in the neutral pion decay (upper panel) and the ALP decay (lower panel), calculated with different methods discussed in Sec.~\ref{sec:HawkingRadiation}. In the pion decay case, the PBH mass is chosen as $M=10^{14.5}~{\rm g}$, corresponding to PBH Hawking temperature $T_H\simeq 33.4~{\rm MeV}$. We also choose PBH abundance to be $f_{\rm PBH}=10^{-9.6}$. In the ALP case, we choose $M=10^{15}~{\rm g}$ with a Hawking temperature of $T_H\simeq 10.6~{\rm MeV}$, $f_{\rm PBH}=10^{-8}$, and $m_a=30~{\rm MeV}$ with an ALP-photon coupling assumed large enough for instantaneous diphoton decay.\footnote{Note in the ALP case, the contribution from neutral pion decay is calculated with the Massless method. We checked the difference between pion decay fluxes in difference scenarios is negligible for the chosen PBH mass in the ALP example since $m_{\pi^0}\gg T_H$.} The rescaled gamma-ray constraints from COMPTEL obtained from \cite{Essig:2013goa} and Fermi observations from \cite{Fermi-LAT:2017opo} for the assumed observation ROI are shown with solid black curves. We also show the future sensitivity of AMEGO-X assuming a $3$-${\rm year}$ all-sky survey~\cite{Fleischhack:2021mhc}.

We compare our massive scalar emission rate calculation from Eq.~\eqref{eq:NRcalculation} (orange, NR calculation) with methods represented with color curves from Eq.~\eqref{eq:massless} (blue, Massless), and from Eq.~\eqref{eq:vsqmassless} (green, Massless$\times v^2$). The orange curve shows the main result of this study. The blue and green curves are obtained with the radiation rate from BlackHawk data for comparison. The upper panel of Fig.~\ref{fig:GammaraySpectrum} shows the case of SM pion production,\footnote{The photon spectra differ between non-relativistic pion production and ALP production from a $10^{14.5}{\rm g}$ PBH. The ALP decay peak is clearly separated from the pion decay signal when $m_a \lesssim 94\,{\rm MeV}$, making it distinguishable based on the photon spectrum. To detect the ALP signal, the photon coupling must be $g_{a\gamma\gamma} \gtrsim 4 \times 10^{-14}{\rm GeV}^{-1} (10\,{\rm MeV}/m_a)^2$ to ensure the decay length is smaller than $r_\odot$. 
} and the lower panel shows the case of a BSM ALP production. The PBH masses are chosen such that the peak location of the Hawking radiation spectrum is close to the scalar particle mass,\footnote{For scalar particle, the peak location of the flux measured at infinity is at $E_{s=0} \simeq 2.81 \, T_H$ \cite{MacGibbon:2007yq}. Note the peak location depends on the greybody factor used in the Hawking radiation spectrum.} in order to show the difference between spectra calculated with different methods. In both examples, the difference in massive scalar particle production rates leads to a different spectrum shape in the energy region to the left of the primary photon peak. In specific, our NR calculation predicts a lower decay photon flux compared to the Massless method, and a higher photon flux than the Massless$\times v^2$ method. One can see the gamma-ray signals are much higher than the statistical uncertainty of future AMEGO-X observation, thus the non-relativistic effect can be tested with the spectrum shape of the observed signal. 

We perform a likelihood analysis on the gamma-ray spectrum to demonstrate the possibility of using AMEGO-X to measure the massive scalar emission rate. The observation time is assumed to be $T_{\rm obs}=3 ~{\rm yrs}$ for the ALP case and $T_{\rm obs}=6 ~{\rm yrs}$ for the pion case. We use the effective area of AMEGO-X measurement reported in \cite{AMEGO:2019gny}. The background photon flux is modeled with the GC component in \cite{Bartels:2017dpb} and cosmic background component in \cite{Caputo:2017tqk}. We do not include Albedo photons from the emission in the Earth’s atmosphere~\cite{Abdo_2009, Harris:2003ce} in our background model, assuming they can be suppressed with a higher mission orbital altitude and selection cuts on the direction of gamma-rays. The photon counts are binned with $45$ bins of equal bin width on logarithmic scale from $5~{\rm MeV}$ to $500~{\rm MeV}$, corresponding to the requirement of about $10\%$ energy resolution. With the above assumptions, we calculate the expected  photon number in the $i{\rm th}$ bin for the true model $n_{i}$ and the test model $\kappa_{i}$. The likelihood that a test model spectrum $\kappa_{i}$ can replicate the true model spectrum $n_{i}$ is defined as
\bea
\mathcal{L}= \exp{\left(\sum_i n_i \ln{(\kappa_i)}-\kappa_i-\ln{(n_i!)}\right)},
\label{eq:likelihood}
\eea
where we assumed the gamma-ray signal is produced following Poisson statistics. We further assume that the joint  analysis of different energy bins follows a $\chi^2$ distribution. The corresponding test statistic ${\rm TS}$ and the observation significance $\sigma$ are obtained as follows~\cite{Cowan:2010js,Rolke:2004mj,Bringmann:2012vr,Fermi-LAT:2015kyq}
\bea
    {\rm TS} = - 2 \ln{\left(\frac{\mathcal{L}}{\mathcal{L}_{\rm true}}\right)} = \sigma^2.
\label{eq:TS}
\eea
Here $\mathcal{L}_{\rm true}$ is the likelihood of the true model calculated with $\kappa_{i}=n_{i}$.

\begin{figure}[ht]
\centering
\includegraphics[width=0.5\linewidth]{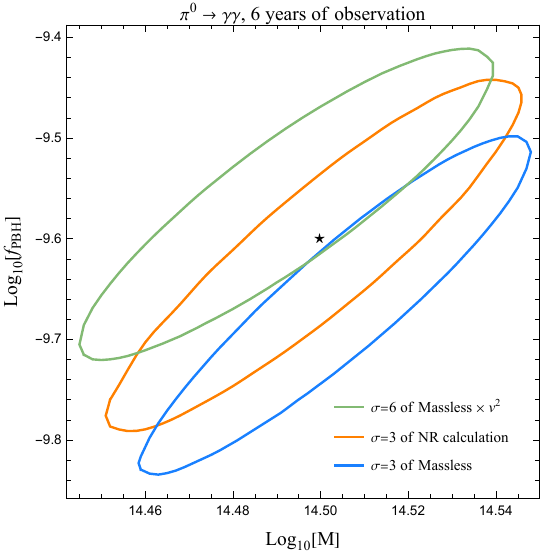}
\\\vspace{1em}
\includegraphics[width=0.51\linewidth]{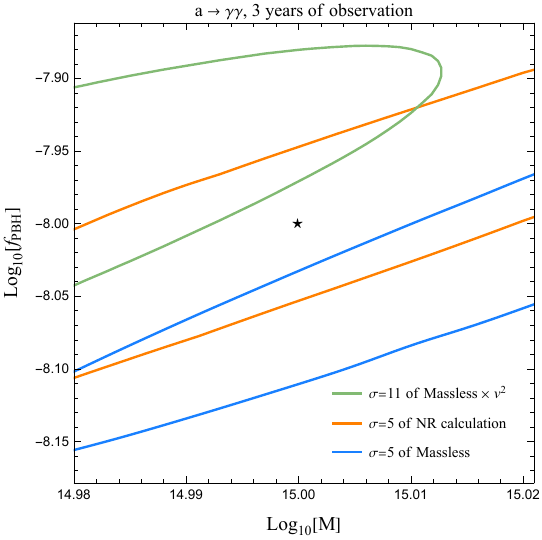}
\caption{Fit to the true model ($\star$) using different assumptions for calculating scalar productions. The significant deviations (in $\sigma$) for the green and blue contours highlight the difficulty of incorporating the correct PBH mass and abundance with incorrect non-relativistic corrections. {\it Upper panel:} The $\pi^0$ case with benchmark values $f_{\rm PBH}=10^{-9.6}$ and $M=10^{14.5}{\rm g}$, assuming 6 years of AMEGO-X observation; {\it Lower panel:} The ALP case with benchmark values $f_{\rm PBH}=10^{-8}$ and $M=10^{15}{\rm g}$, assuming 3 years of observation. The tidal force effect on $\pi^0$ production is assumed negligible. The nearly parallel curves come from cutting the elliptical contours for the PBH mass window in the plot.}
\label{fig:significance}
\end{figure}

In Fig.~\ref{fig:significance}, we demonstrate that the observed gamma-ray signal can distinguish the non-relativistic effect between our method and approximations Eqs.~(\ref{eq:massless}) and~(\ref{eq:vsqmassless}) used in the literature. We start by selecting true model PBH parameters $\{M, f_{\rm PBH}\}$ (marked by black stars "$\star$") and calculate the corresponding photon spectrum $n_i$ using Eq.~\eqref{eq:NRcalculation}. We then scan test models using NR calculation, Massless, and Massless$\times v^2$ calculations to obtain $\sigma^2$ for various $\{M, f_{\rm PBH}\}$ values. Best-fit regions are shown with color contours indicating specific $\sigma$ values.

The orange contours, based on the correct non-relativistic calculation, are centered around the true model point. Regions within these contours have smaller significance in $\sigma$ than the labeled values, making the test models indistinguishable within AMEGO-X sensitivity. Due to incorrect mass corrections, the best-fit points in the green and blue contours shift away from the black stars, and these calculations disfavor the true $\{M, f_{\rm PBH}\}$ values with similar or greater significance indicated by the contours.

The upper panel of Fig.~\ref{fig:significance} shows the likelihood fit in the case of $\pi^0$ production, assuming 6 years of the AMEGO-X observation. In this study, we do not consider the tidal force effect discussed in Sec.~\ref{sec.tidle} for the pion production. The true PBH model we use is $\{M,f_{\rm PBH}\}=\{10^{14.5}~{\rm g}, 10^{-9.6}\}$, shown with the black star in the upper panel. The fit using the NR calculation (orange) can successfully locate the true model point as the black star is well inside the orange contour. In contrast, methods without the correct non-relativistic correction fail to accurately fit the true PBH parameters. The true model lies outside the $3\sigma$ contour for the Massless calculation (blue) and is only marginally within the $6\sigma$ contour for the Massless$\times v^2$ calculation (green). This demonstrates the importance of correct mass correction in $\pi^0$ production for determining PBH parameters from the AMEGO-X search.

The lower panel of Fig.~\ref{fig:significance} shows a similar comparison between different methods for the ALP production, assuming 3 years of the AMEGO-X observation. The true PBH parameters are $\{M,f_{\rm PBH}\}=\{10^{15}~{\rm g}, 10^{-8.0}\}$. Although the mass of the ALP is a priori unknown, we fix $m_a = 30~{\rm MeV}$ for simplicity in the parameter scan. We find that the true model point lies well within the orange contour. The true PBH parameters lie outside the $5\sigma$ contour for the Massless calculation and the $11\sigma$ contour for the Massless$\times v^2$ calculation, showing a significant deviation between the gamma-ray spectra obtained with correct and incorrect non-relativistic corrections.

Notice that the blue contours in the plots are generally below the orange contours because the production rate of massless particles is higher than that of massive ones with the same PBH mass and abundance, leading to smaller predicted $f_{\rm PBH}$ values. Conversely, the Massless$\times v^2$ method (green) predicts an overly small ALP production rate, placing the green contour in regions with larger $f_{\rm PBH}$ values.

\section{Discussion and Conclusion}
\label{sec:discussion}
In this study, we reassess the calculation of Hawking radiation for massive scalar particles~\href{https://github.com/Haoran-Brook/HoRNS}{\faGithub} and extend our analysis to include their impact on $\pi^0$ and ALP production, influencing secondary gamma-ray emissions. Our calculation reveals significant non-relativistic corrections in scalar production signals from black holes of approximately $10^{15}$~g when compared to the sensitivity of next-generation gamma-ray detectors like AMEGO-X. Even in the absence of beyond the SM particles, inaccurately incorporating the $\pi^0$ mass can substantially impair signal fitting. As discussed in~\cite{Agashe:2022jgk}, if PBHs originate from the collapse of large primordial curvature perturbations, the observable $f_{\rm PBH}$ in next-generation gamma-ray detectors will correspond to substantial curvature perturbations, yielding significant gravitational wave signals detectable by future gravitational wave detectors. Even if the PBH mass spectrum is not monochromatic as assumed in this work, the interplay between gravitational wave and gamma-ray detections can provide good measurements of the PBH spectrum. Accurate predictions of Hawking radiation signals, encompassing the non-relativistic particle productions discussed herein, will be crucial for identifying PBH properties via such multi-messenger measurements. Although the current work is focused on charge-neutral massive scalar particles and Schwarzschild black holes, a similar analysis can be performed for general conditions. 

As noted earlier, tidal forces can significantly impact $\pi^0$ production for black hole masses relevant to AMEGO-X measurements. If tidal forces strongly suppress pion production, ALPs with masses close to the pion's could mimic the expected pion signal without disruption. A better understanding of hadron production via Hawking radiation is crucial for accurately determining PBH properties from its emissions.

The observed photon spectrum also depends on other black hole properties, such as spin, charge, and mass variation during evaporation. The total gamma-ray spectra can be obtained by integrating these effects over time. Additionally, if a PBH and its emitted particles share the same charge sign, their electromagnetic interaction can enhance production rates, impacting the photon spectrum~\cite{Page1977}. For rotating black holes, emissions are influenced by angular momentum alignment with the black hole's spin~\cite{Page:1976df}. While these factors modify the gamma-ray spectra, they complement the non-relativistic corrections discussed here. Non-relativistic effects become relevant if heavier particles, like hadronic states, are produced near their mass threshold. A comprehensive analysis of all these factors is left for future work.

\acknowledgments
We thank Ragnar Stroberg for the helpful discussions. We thank Jane H. MacGibbon for the helpful feedback on the existing literature regarding massive scalar production. The research of YT is supported by the National Science Foundation Grant Number PHY-2112540. The work of TX is supported by the U.S.~Department of Energy Grant DE-SC0009956. YT would like to
thank the Tom and Carolyn Marquez Chair fund for its generous support. YT would also like to thank the Aspen Center for Physics (supported by NSF grant PHY-2210452) and Kavli Institute for Theoretical Physics (KITP, supported by grant NSF PHY-2309135) for their hospitality while this work was in progress. 

\appendix

\section{Transition and Reflection Amplitudes}
\label{amplitude justification}
In this section, we demonstrate how to identify the physical meaning of the amplitudes in the asymptotic solution in Eq.~(\ref{eq.R}). 

First, one may assume a more general form for the asymptotic solutions, 
\begin{align}
R(l, r,\omega, M)=
\left\{
\begin{array}{cc}
   A e^{-i\omega r^*},  &  r_s=2GM\\
    B  e^{-i\omega v r^*}+ C e^{i\omega v r^*}, & r^*\to r\to \infty,
\end{array}
    \right.
    \label{general R}
\end{align}
where $A, B$ and $C$ are unknown amplitudes. Then introducing the Wronskian operator, 
\begin{align}
    W[R,R^*]:=R\frac{\partial R^*}{\partial r^*}-R^*\frac{\partial R}{\partial r^*},
\label{wronskian definition}
\end{align}
where $R$ and $R^*$ are the radial wave function and its complex conjugate defined by Eqs.~(\ref{eq.R}) and (\ref{radial eq}), and $r^*$ is the tortoise radial coordinate introduced before, $\frac{d}{d r^*}=f(r)\frac{d}{dr}$. By the fact that Eq.~(\ref{radial eq}) has no first-order derivative in terms of $r^*$, it can be proved that 
\[
\frac{\partial W[R,R^*]}{\partial r^*}=0,
\] so $W[R, R^*]$ is a constant along $r^{*}$, which is related to the conservation of the flux. Inserting the general solution Eq.~(\ref{general R}) to the Wronskian, 
\begin{align}
W[R,R^*]=
\left\{
\begin{array}{cc}
  -2i\omega |A|^2 ,  &  r_s=2GM\\
    -2i \omega v |B|^2 +2i\omega v |C|^2, & r^*\to r\to \infty.
\end{array}
    \right.
    \label{wronskian}
\end{align}
Therefore,
\begin{align}
    -2i\omega |A|^2=&-2i \omega v |B|^2 +2i\omega v |C|^2\nonumber\\
    \Rightarrow \frac{|A|^2}{v}&+|C|^2=|B|^2.
\end{align}

Physically, $\frac{|C|^2}{|B|^2}$ should be the reflection rate defined at infinity, so we may normalize $B$ as 1 and set $C$ as the reflection amplitude $R_{\omega l}$ correspondingly. Therefore, $\frac{|A|^2}{v}$ should be the transition rate $|T_{\omega l}|^2$. Hence the ansatz Eq.~(\ref{eq.R}).

\section{Hawking Radiation and Penrose Diagram}
\label{Penrose}

In this section, we describe the Hawking radiation and time-reversed propagation using the Penrose diagram~\cite{Misner:1973prb}, which compresses spacetime into a finite region via conformal transformation. We focus on providing a concrete picture of these processes without delving into mathematical details, discussing only the Schwarzschild-type spacetime relevant to our study.

The original Schwarzschild metric
\[
    ds^2=f(r) dt^2- \frac{1}{f(r)}dr^2-r^2d\theta^2-r^2\text{sin}\theta^2 d\phi^2
\]
covers only the region outside the black hole and extends to spatial and temporal infinities. To show the whole spacetime in a finite size, the following new conformal coordinates $(T,R)$ are introduced, 
\begin{align}
    \tan (R-T)&=e^{\frac{r^*-t}{4GM}},\nonumber\\
    \tan (R+T)&=e^{\frac{r^*+t}{4GM}},
\end{align}
where $r^*=r+2GM \ln (\frac{r}{2GM}-1)$ is the tortoise coordinate. The new coordinates are referred to as conformal time T and conformal radius R. The special property that $\tan (\pm\frac{\pi}{2})\to\pm\infty$ allows us to show the points at infinity. The Penrose diagram of a Schwarzschild black hole is shown below (without the white hole interior and another exterior region). 

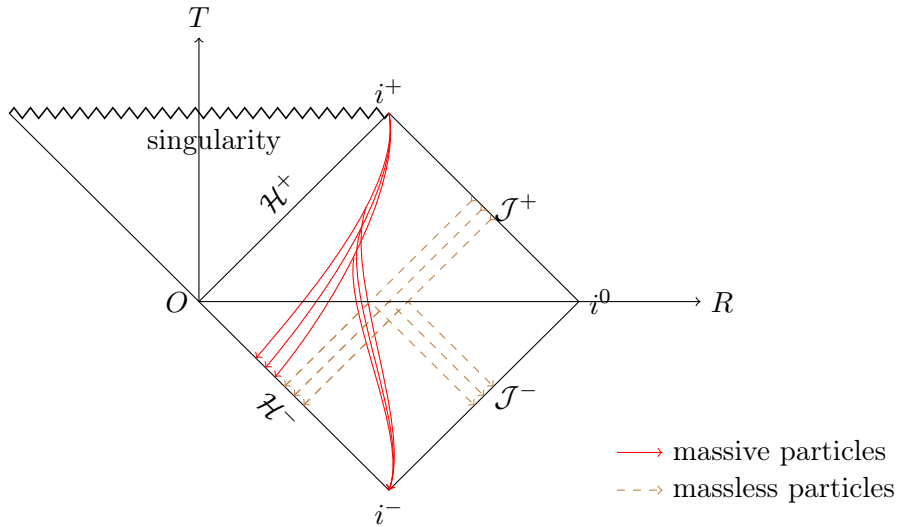
\begin{figure}[ht]
\centering
\begin{tikzpicture}
\node (I)    at ( -2.5,0)   {};
\path  
  (I) +(90:2.5)  coordinate[label=90:$i^+$]  (Itop)
       +(-90:2.5) coordinate[label=-90:$i^-$] (Ibot)
       +(0:2.5)   coordinate[label=0:$i^0$]                  (Iright)
       +(180:2.5) coordinate[label=180:$O$] (Ileft)
       ;
\draw[singularity] (Itop) -- node[pos=0.46,below] {\strut singularity} (-7.5,2.5);
\draw (Iright) -- 
          node[midway, above, right]    {$\cal{J}^+$}
      (Itop) --
          node[midway, above, sloped] {$\mathcal{H}^+$}
      (Ileft) -- 
          node[midway, below, sloped] {$\mathcal{H}^-$}
      (Ibot) --
          node[midway, below, right]    {$\cal{J}^-$}
      (Iright);
\draw 
      (Ileft) -- 
          node[midway, above, sloped]{} 
    (-7.5, 2.5);
\draw[<-][red]
    ($(Ileft)!.3!(Ibot)$) to[out=50, in=-80, looseness=0.7] (Itop);
\draw[<-][red]
 ($(Ileft)!.35!(Ibot)$) to[out=50, in=-80, looseness=0.7] (Itop);
\draw[<-][red]
    ($(Ileft)!.4!(Ibot)$) to[out=50, in=-80, looseness=0.7] (Itop);
\draw[->][red]
    (-2.8,1.25) to[out=-110, in=70, looseness=0.63]  (Ibot);
\draw[->] [red]
    (-2.87,0.95) to[out=-110, in=70, looseness=0.63]  (Ibot);
\draw[->][red]
    (-2.96,0.585) to[out=-100, in=70, looseness=0.63]  (Ibot);
 \draw[dashed][<-<][brown]
    ($(Ileft)!.45!(Ibot)$) -- ($(Itop)!0.45!(Iright)$);
\draw[dashed, <-<, brown]
  ($(Ileft)!.5!(Ibot)$) -- ($(Itop)!.5!(Iright)$);
\draw[dashed, <-<, brown]
  ($(Ileft)!.55!(Ibot)$) -- ($(Itop)!.55!(Iright)$);
\draw[dashed, ->, brown]
  ($(Ileft)!.45!(Ibot)$) -- (-180:2.75) -- ($(Iright)!0.55!(Ibot)$);
\draw[dashed, ->, brown]
  ($(Ileft)!.5!(Ibot)$) -- (-180:2.5) -- ($(Iright)!.5!(Ibot)$);
\draw[dashed, ->, brown]
 ($(Ileft)!.55!(Ibot)$) -- (-180:2.25) -- ($(Iright)!.45!(Ibot)$);
 \draw[->] (Ileft) -- (1.6,0) node[right] {$R$};
\draw[->] (Ileft) -- (-5,3.5) node[above] {$T$};
\draw[->, red]
(0.5,-2) -- (1.1,-2) node[right, black] {massive particles};
 \draw[dashed,->,brown]
 (0.5,-2.5) -- (1.1,-2.5) node[right][black] {massless particles};
\end{tikzpicture}
    \caption{A fraction of the Penrose diagram of a fully extended Schwarzschild black hole, representing the time-reversed propagation of massive fields from $i^+$ and massless fields from $\cal{J}^+$. Each point in the diagram corresponds to a sphere $S^2$ with a radius determined by its coordinate.}
    \label{penrose}
\end{figure}

The meaning of the labels on the diagram is the following,
\begin{itemize}
    \item when $r^{*}=cons.$ and $t\to\pm\infty$, one approaches to $R\to\frac{\pi}{4}$, $T\to\pm\frac{\pi}{4}$, therefore the two points on the $(T,R)$ plane $i^{\pm}=(\frac{\pi}{4}, \pm\frac{\pi}{4})$ represent the time-like future/past infinity, at where all world-lines of massive particles converge (if not going into the black hole);
    \item when $r\to+\infty, r^{*}\to+\infty$, and$t=cons.$, one approaches to $R\to\frac{\pi}{2}$ and $T\to 0$, therefore  $i^0=(\frac{\pi}{2},0)$ represents the space-like infinity;
    \item when $r^{*}\pm \,t=cons.$, we have $R\pm T=cons'.$, so all 45$^\circ$ lines represent null trajectories. Especially, $\cal{H}^{\pm}$ lines are approached when $r=2GM$, $r^{*}\to-\infty$, and $t\to\pm\infty$, so they represent the future and past black hole horizons. The junction $O$ of $\cal{H}^-$ and $\cal{H}^+$ is the bifurcation surface. The left region of the future horizon $\cal{H}^+$ is the black hole interior, while the left region of the past horizon $\cal{H}^-$ is sometimes called the white hole interior (not included). $\cal{J}^{\pm}$ are approached when $r\to\infty, t\to\pm\infty$, so they are the future and past null infinity respectively, at where massless particles converge (if not going into the black hole);
    \item when $r=0$, one hits the singularity. When $t=0$, $T=0$, and $\tan R=e^{r^{*}/4GM}$.
\end{itemize}

The Hawking radiation starts from some point outside $\mathcal{H}^{+}$ (effectively a tunneling process \cite{Parikh:1999mf} from the black hole interior to some point outside) and reaches $i^+$ (massive) or $\mathcal{J}^+$ (massless). The greybody factor $\Gamma$ represents the transition rate from $\mathcal{H}^+$ to $i^+$ or $\cal{J}^+$.  If the radiation is propagating backward in time, the penetration rate then connects $i^+$ (or $\cal{J}^+$) to $\mathcal{H}^-$. Ignoring the backreactions of Hawking radiation, the effective gravitational barrier can be treated as time-invariant, therefore the transition rate $\Gamma$ is the same and we use the time-reversed process to calculate $\Gamma$. 

The bifurcation of each curve on the diagram indicates that a fraction $1-\Gamma$ is scattered away and a fraction $\Gamma$ penetrates the gravity barrier. However, the precise location of bifurcations can not be determined precisely due to the spread of the effective gravity potential and the waves so the locations are only for illustrative purpose. 

\bibliography{reference}
\bibliographystyle{JHEP}

\end{document}